\documentclass[iop]{emulateapj}

\usepackage{amssymb}
\usepackage{amsmath}	
\usepackage{multirow}

\shorttitle{Obscured Galactic Giant H\,{\sc{ii}} Regions. VII.}
\shortauthors{Navarete et al.}

\begin{document}
	
\title{The Stellar Content of Obscured Galactic Giant H\,{\sc{ii}} Regions. VII. W3}

\author{F. Navarete\altaffilmark{1}, E. Figueredo\altaffilmark{1,2}, A. Damineli\altaffilmark{1}, A.~P. Mois\'{e}s\altaffilmark{1,3}, R. D. Blum\altaffilmark{4} and P. S. Conti\altaffilmark{5}}

\altaffiltext{1}{Instituto de Astronomia, Geof\'{i}sica e Ci\^{e}ncias Atmosf\'{e}ricas, Universidade de S\~{a}o Paulo, R. do Mat\~ao, 1226, 05508-090, S\~ao Paulo, SP, Brazil; navarete@usp.br}
\altaffiltext{2}{The Open University, UK.}
\altaffiltext{3}{UNIVASF, Rua Jo\~{a}o Ferreira dos Santos, 64770-000, S\~{a}o Raimundo Nonato, PI, Brazill}
\altaffiltext{4}{NOAO, 950 N Cherry Ave., Tuczon, AZ 85719 USA.}
\altaffiltext{5}{JILA, University of Colorado, Boulder, CO 80309-0440, USA.}

\begin{abstract}

Spectrophotometric distances in the $K \ $band have been reported by different authors for a number of obscured Galactic  H\,{\sc{ii}} regions. Almost 50\% of them show large discrepancies compared to the classical method using radial velocities measured in the radio spectral region. In order to provide a crucial test of both methods, we selected a target which does not present particular difficulty for any method and which has been measured by as many techniques as possible.  The W3 star forming complex, located in the Perseus arm, offers a splendid opportunity for such a task.  We used the NIFS spectrograph on the Frederick C. Gillett Gemini North telescope to classify candidate ``naked photosphere'' OB stars based on 2MASS photometry. Two of the targets are revealed to be mid O-type main sequence stars leading to a distance of d = 2.20 kpc. This is in excellent agreement with the spectrophotometric distance derived in the optical band \citep[d = 2.18 kpc,][]{Humphreys1978} and with a measurement of the W3 trigonometric parallax \citep[d = 1.95 kpc,][]{Xu2006}. Such results confirm that the spectrophotometric distances in the $K \ $band are reliable. The radio derived kinematic distance, on the contrary, gives a distance twice as large \citep[d = 4.2 kpc,][]{Russeil2003}.  This indicates that this region of Perseus arm does not follow the Galactic rotation curve, and this may be the case also for other H\,{\sc{ii}} regions for which discrepancies have been found. 

\end{abstract}



\section{Introduction} \label{introduction}

Giant H\,{\sc{ii}} regions ($N_{LyC}$ $>$ $10^{50}$ photons per second) are the best tracers of the spiral structure of the Milky Way, and the usual procedure to obtain their heliocentric distances is by using longer wavelengths to overcome the optical obscuration from interstellar extinction. 
Radial velocities from radio recombination lines connected with a rotation model for the Galaxy is the classical tool to derive kinematic distances. A large catalogue was published by \citet[][]{Russeil2003}, which is presently the main reference to draw the position of the spiral arms and to calculate the luminosity in the Lyman continuum, from which the star formation rates throughout the Galaxy are derived. Recent work has shown discrepancies with this method, showing that our picture of the Galaxy, based on the kinematic method, may not be accurate.

Other methodologies that do not use galactic rotation models (non-kinematic methodologies), such as trigonometric and spectrophotometric parallaxes, have given different results for the distances to these star forming regions when compared to the kinematic ones. 

Trigonometric distances, such as those derived from radio VLBI, are very accurate for relatively close targets, since they are based on simple triangulation from the Earth's orbit. The Gaia survey will extended the usage of such methodology to farther targets, since it will achieve a spatial resolution better than a factor of 1000 compared to Hipparcos, reaching microarcsec precision.\citet{Imai2000,Chen2006,Reid2009a,Reid2009b,Reid2009c,Moscadelli2009,Xu2009,Xu2006,
Zhang2009,Brunthaler2009,Bartkiewicz2008,Moellenbrock2009,Sato2008,Hachisuka2006,Hachisuka2009,Honma2007,Hirota2007,
Menten2007} and \citet{Choi2008} reported trigonometric distances for 19 sources, from which eight (42\%) are substantially closer and one farther than published kinematic distances. Unfortunately, the size of the Earth's orbit is a limiting factor for this method, which cannot be applied to the whole Galaxy. 

Spectrophotometric distances using near infrared (NIR) photometry and $K \ $band spectroscopy have been reported by 
\citet{Hanson1996,Blum1999,Blum2000,Blum2001,Figueredo2002,Bik2005,Figueredo2005} and \citet{Figueredo2008}. \citet{Moises2011} reviewed and revised the spectrophotometric distances. All the distances were referred to a common Galactic center -- sun distance (R$_0$ = 8.5 kpc) and extinction law. Since there are different reddening laws in the literature, \citet{Moises2011} derived the extremes and adopted the mean value, resulting in a mean distance. The range between the extremes was adopted as the uncertainty. The adopted laws were taken from the work of \citet{Mathis1990} and \citet{Stead2009}. The major source of uncertainty in distance by the spectrophotometric technique, however, is given by the scatter in the calibration of the absolute magnitude of O-type stars ($\Delta$K = 0.67 which translates to an error of 30\% over the obtained distance). From 17 sources, eight (~50\%) display spectrophotometric distances smaller than kinematic and one a larger distance.

In this paper, we present a NIR study of the H\,{\sc{ii}} region W3, which is located in the Perseus spiral arm (R.A. = 2$^h$26$^m$34$^s$, Decl. = 62$^{\circ}$00'43''). W3 offers an excellent opportunity to discuss the distinct methods used to measure distances. Indeed, the W3 distances were derived by several authors, using methodologies other than the kinematic one. The near infrared study presented here, comprises $J$, $H$ and $K_{s}-$band photometry of this complex of H\,{\sc{ii}} regions and the W3 $K-$band spectrophotometric distance. 

The problem with kinematic distances can be underlined by considering the work of \citet{Russeil2003}, who derived a kinematic distance to the W3 complex of 4.2 kpc. Non-kinematic determinations (that do not use Galactic rotation models) have obtained smaller distances as follows. 
By measuring the annual parallax of the H$_{2}$O masers, \citet{Moellenbrock2009} derived a distance to the W3 infrared source IRAS 00420+5530 of $2.17$ $\pm$ $0.05$ kpc. 
\citet{Hachisuka2006} and \citet{Chen2006} found a distance of $2.04$ $\pm$ $0.07$ kpc, which is consistent with the W3(OH) distance derived by \citet{Xu2006}, who studied the trigonometric parallax of CH$_{3}$OH maser and derived a distance of 1.95 $\pm$ 0.04 kpc.
Using H$_{2}$O masers and their radial velocities and proper motions associated with two different outflows sources, \citet{Imai2000} obtained a distance to the W3 infrared source IRS 5 d = $1.9$ $\pm$ $0.3$ kpc.  

All these trigonometric distances disagree with the kinematic results, roughly, by the same amount, about half of the kinematic value. They are in agreement with the distance derived by \citet{Humphreys1978}, $d$ = 2.18 kpc, using spectrophotometric parallax in the optical domain for a large number of OB stars in the direction of the Perseus arm.

Using photometry extracted from the 2MASS catalogue, we have constructed color-color (C-C) and color-magnitude (C-M) diagrams and selected seven candidate ``naked photosphere'' OB stars to follow up with $ K \ $band spectroscopy. 

This work is organized as follows. In \S 2, the observations and data reduction are described. In \S3 we present the main results from the investigation of the W3 complex through {\it J-}, {\it H-} and $K_{s}-$band photometry and $ K \ $band spectroscopy of the ionizing sources. \S4 details the distance determination to the W3 complex. Finally, the overall discussion of the results and the usual procedure to obtain their heliocentric distances are given in \S5.


\section{Observations and Data Reduction} \label{observations}

\subsection{Observations}

J ($\lambda \thicksim 1.24 \ \mu m$), H ($\lambda \thicksim 1.64 \ \mu m$) and K$_{s}$ ($\lambda \thicksim 2.22 \ \mu m$) images of the W3 complex were obtained from the 2MASS\footnote{http://www.ipac.caltech.edu/2mass/} survey \citep{Skrutskie2006}. 
We have also used IRAC-Spitzer images in 8.0 $\mu$m (red), 5.8 $\mu$m (green) and 3.6 $\mu$m (blue) channels \citep{Fazio2004}. The images were obtained using the software Leopard\footnote{http://archive.spitzer.caltech.edu/} and are from the Spitzer program ID number 127.

$K$-band ($\lambda$ $\approx$ 2.2 $\mu$m) spectra were obtained with the NIFS\footnote{the Near-infrared Integral-Field Spectrograph was built by the Research School of Astronomy and Astrophysics which is part of the Institute of Advanced Studies of the Australian National University. See \citet{McGregor2003} for a description of NIFS.} spectrograph at the Frederick C. Gillett Gemini North $8~m$ telescope at Mauna Kea, Hawaii, on the nights of 2008 September 24th, October 14th and the nights of December 10th, 17th to 21st and 24th. Data were taken in queue-scheduled mode (GN-2008B-Q92). NIFS was used with the facility adaptive optics module (ALTAIR) in natural guide stars (NGS) mode. 
NIFS has a full wavelength coverage at {\it K} of about 4200~\AA \ and a linear dispersion of 2.13~\AA \ pixel$^{-1}$ resulting in a spectral resolving power of $\lambda/\Delta\lambda$ = 5160 in the {\it K}-band \citep{Blum2008}. 

For each observation, three frames were taken on source and three sky frames were obtained by pointing the telescope on a nearby ($\approx$ 30\arcsec) blank field. After each exposure, the telescope was offset by a few arcsecs (between 5\arcsec and 45\arcsec) in order to avoid cosmetic defects when combining the 2D spectra. Each frame had the same exposure time.

The star HD~15086 was used as telluric reference for all targets. This object was observed with a total exposure time of 60 seconds, split over four frames.  

The air mass of the observations ranged between 1.35 and 1.46. The difference between science objects and the telluric star airmass was less than 0.2 in all cases and no correction was applied. 

\subsection{Data Reduction} \label{reduction}

Data reduction was performed using the Gemini NIFS IRAF\footnote{IRAF is distributed by the National Optical Astronomy Observatory, which is operated by the Association of Universities for Research in Astronomy, Inc., under cooperative agreement with the National Science Foundation.} package.
Calibration data were prepared with Gemini-IRAF NFPREPARE task which adds the Data Quality and Variance extensions to each file.  Flat-field frames were created from quartz lamp spectra. The wavelength transformation was obtained using {\it K-}band telluric lines from the sky spectrum, obtained after each observation. The spatial correction was made using the Ronchi mask \citep[for more information, see][]{Blum2008}.
After the baseline procedure, data from the telluric star were prepared with NFPREPARE, were flat-fielded, wavelength corrected and sky subtracted. The sky frames were combined using a median image of four dithered frames.
Wavelength transformation was accomplished by using the sky spectrum, obtained as described previously, and one-dimensional spectra were then extracted from a 0".5 radius circle centered over the telluric star in the FITS data cube.
Finally, the Br$\gamma$ feature was removed by fitting a Voigt profile between two continuum points on the HD~15086 spectrum. 

Science data were reduced with the same procedure as for the telluric star. The resulting object spectrum was divided by the telluric star spectrum. One-dimensional spectra were then extracted from a 0.3'' radius circle centered on each target in the FITS Data cube.
The science spectra were subtracted by a nebular spectrum, extracted over an annular aperture of 0.6'' external radius and 0.3'' inner radius, to remove non-photospheric features (such as Br-$\gamma$ contamination over the spectra of B-type stars). Before subtraction, we scaled the sky annulus and the circle around the star by their respective area on the image.
The spatial resolution of our data ranged from 0.18'' to 0.67'' FWHM, measured at 2.107 $\mu$m on each spectrum.

The $K$-band classifications from the \citet{Hanson1996,Hanson2005} catalogue were adopted to classify the spectral types.
The spectral lines used were the C\,{\sc{iv}} triplet at 2.069 $\mu$m, 2.078 $\mu$m and 2.083 $\mu$m, He\,{\sc{i}} at 2.1126 $\mu$m and 2.1625 $\mu$m, N\,{\sc{iii}} at 2.1155 $\mu$m and He\,{\sc{ii}} at 2.1885 $\mu$m.
Young stars in H\,{\sc{ii}} regions usually have their spectra contaminated by the 2.058 $\mu$m He\,{\sc{i}} and Br$\gamma$ nebular emission. However, the classification of O-type stars does not require these lines \citep{Hanson1996}.
On the other hand, the classification of B stars is complicated by the absence of features sensitive to stellar temperature and spectral type (C\,{\sc{iv}},  N\,{\sc{iii}}, and He\,{\sc{ii}}). The only signature of these stars are the Br$\gamma$ and He\,{\sc{i}} features. This classification scheme is useful for determining the stellar spectral type, but is not a precise tool for luminosity-class indicators. The presence of a H\,{\sc{ii}} region, however, ensures that the stars are young, close to the main sequence.

\section{Results} \label{results}

\subsection{Photometry} \label{resim}

\begin{figure*}[!ht]

	\begin{minipage}[b]{0.49\linewidth}
		\includegraphics[width=\linewidth]{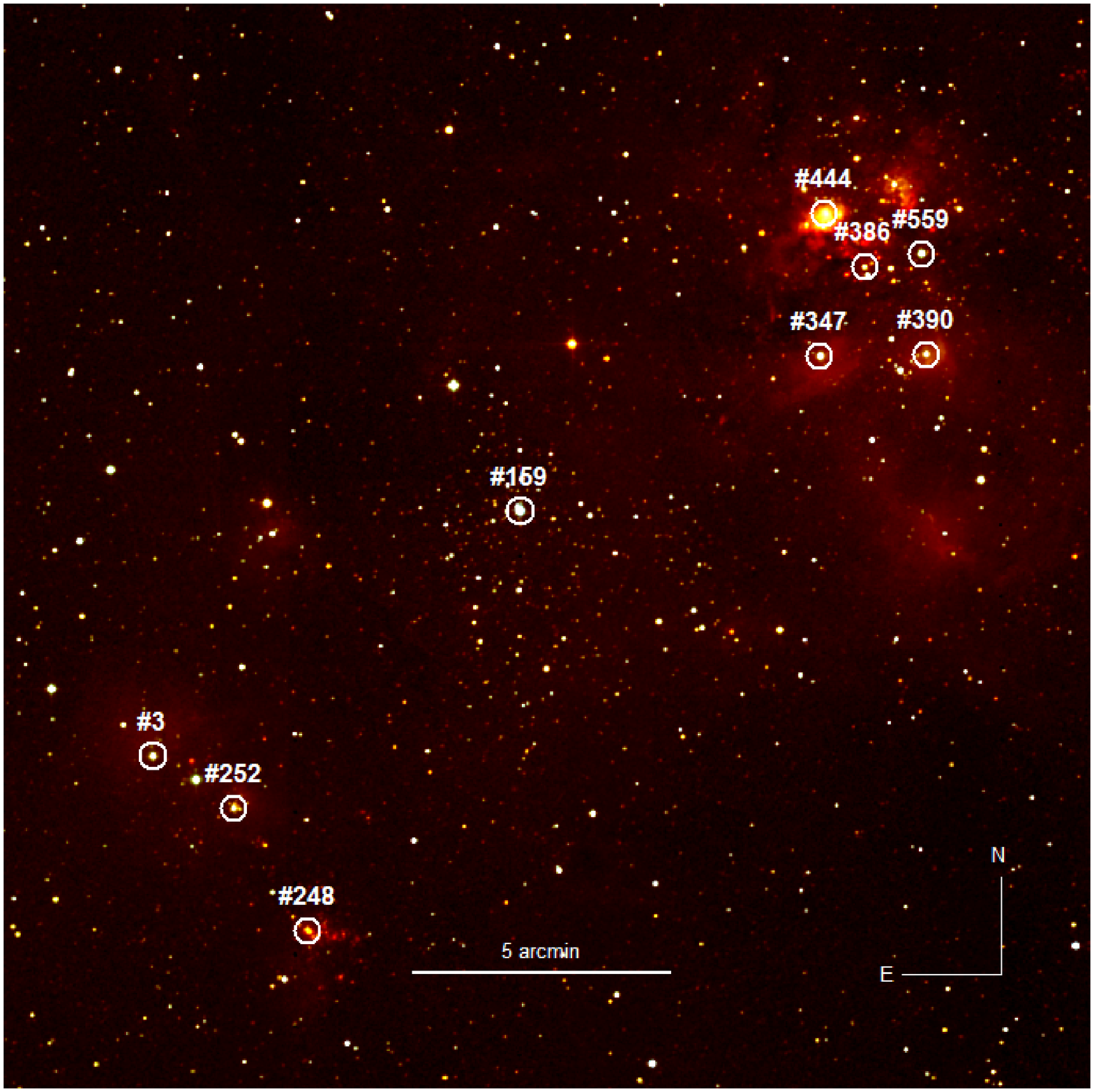}
	\end{minipage}
	\begin{minipage}[b]{0.49\linewidth}
		\includegraphics[width=\linewidth]{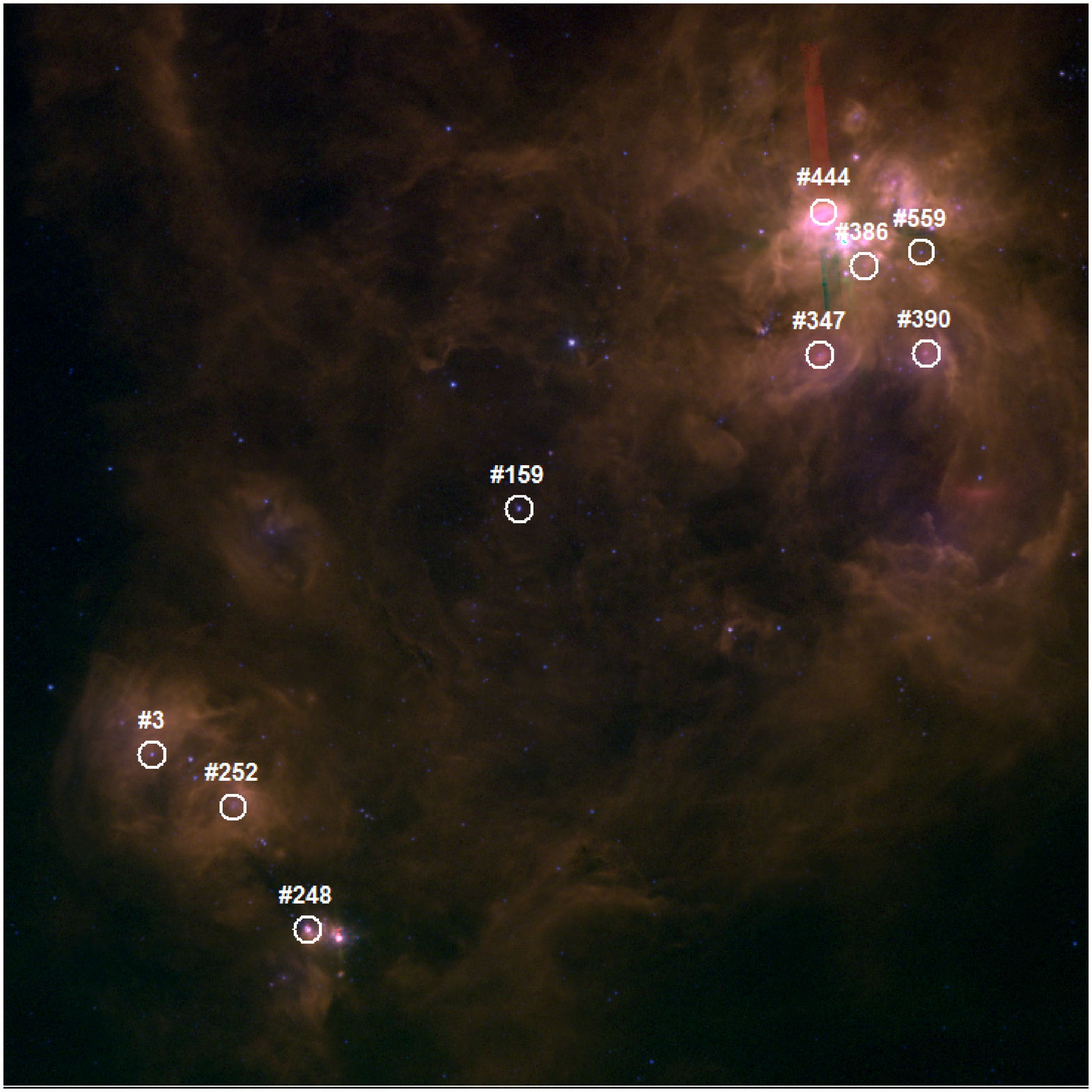}
	\end{minipage}

\caption{{Left panel: 2MASS false-color image of the W3 complex: $J$ is blue, $H$ is green and $K_{s}$ is red. Right panel: Spitzer false-color image of the same field of view: 3.6 $\mu$m is blue, 5.8 $\mu$m is green and 8.0 $\mu$m is red. Both images have a $20' \times 20'$ field-of-view, centered on R.A. = 2$^{h}$26$^{m}$34$^{s}$.0, Decl. = 62$^{\circ}$00\arcmin43\arcsec.3 (J2000). North is up and east is to the left.}}
\label{fig:images-color}
\end{figure*}

Figure~\ref{fig:images-color}a shows a false-color image, made by a composition of three near-infrared 2MASS images $J$ (blue), $H$ (green) and $K_{s}$ (red) of the W3 complex. As can be seen, the observed objects (labbeled) are distributed about the W3 complex, allowing a distance estimation of the entire structure. In this image, a cluster of stars in the center of the field can be seen and also two regions with strong nebular emission can be seen to the NW and to the SE, surrounding the central cluster. Stars discussed below and observed spectroscopically are indicated in the figure.

The Spitzer image (Figure~\ref{fig:images-color}b) shows the nebulosity of this field in detail and reveals that these nebular regions are connected and form a larger and single structure, which cannot be concluded by just analysing the 2MASS image. Also, the Spitzer image reveals that the central cluster of the W3 complex shows weak nebular emission when compared to other structures on the same field. Stars discussed below and observed spectroscopically are indicated in the figure.

\begin{figure*}[!ht]

	\begin{minipage}[b]{0.505\linewidth}
		\includegraphics[width=\linewidth]{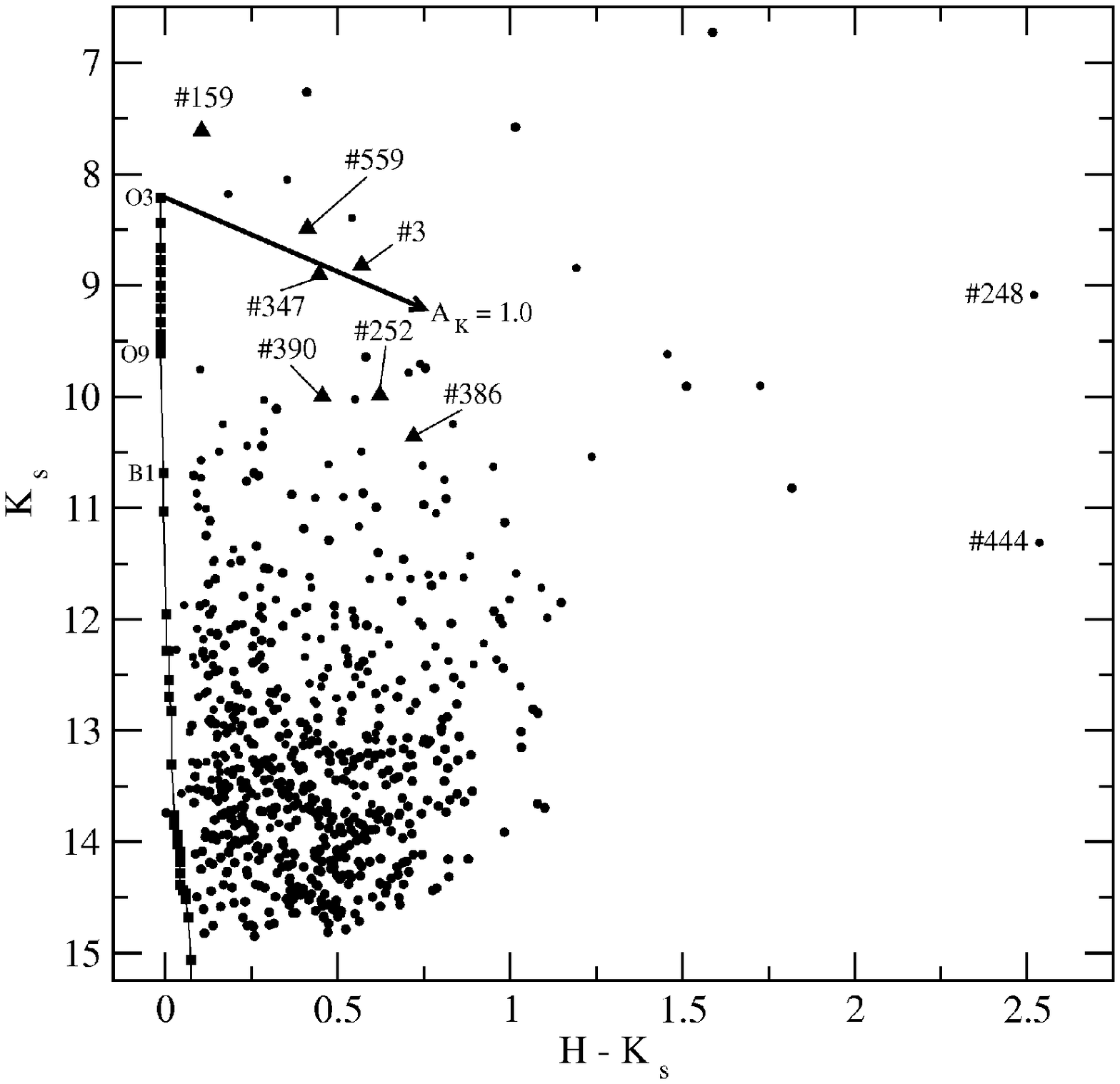}
	\end{minipage}
	\begin{minipage}[b]{0.48\linewidth}
		\includegraphics[width=\linewidth]{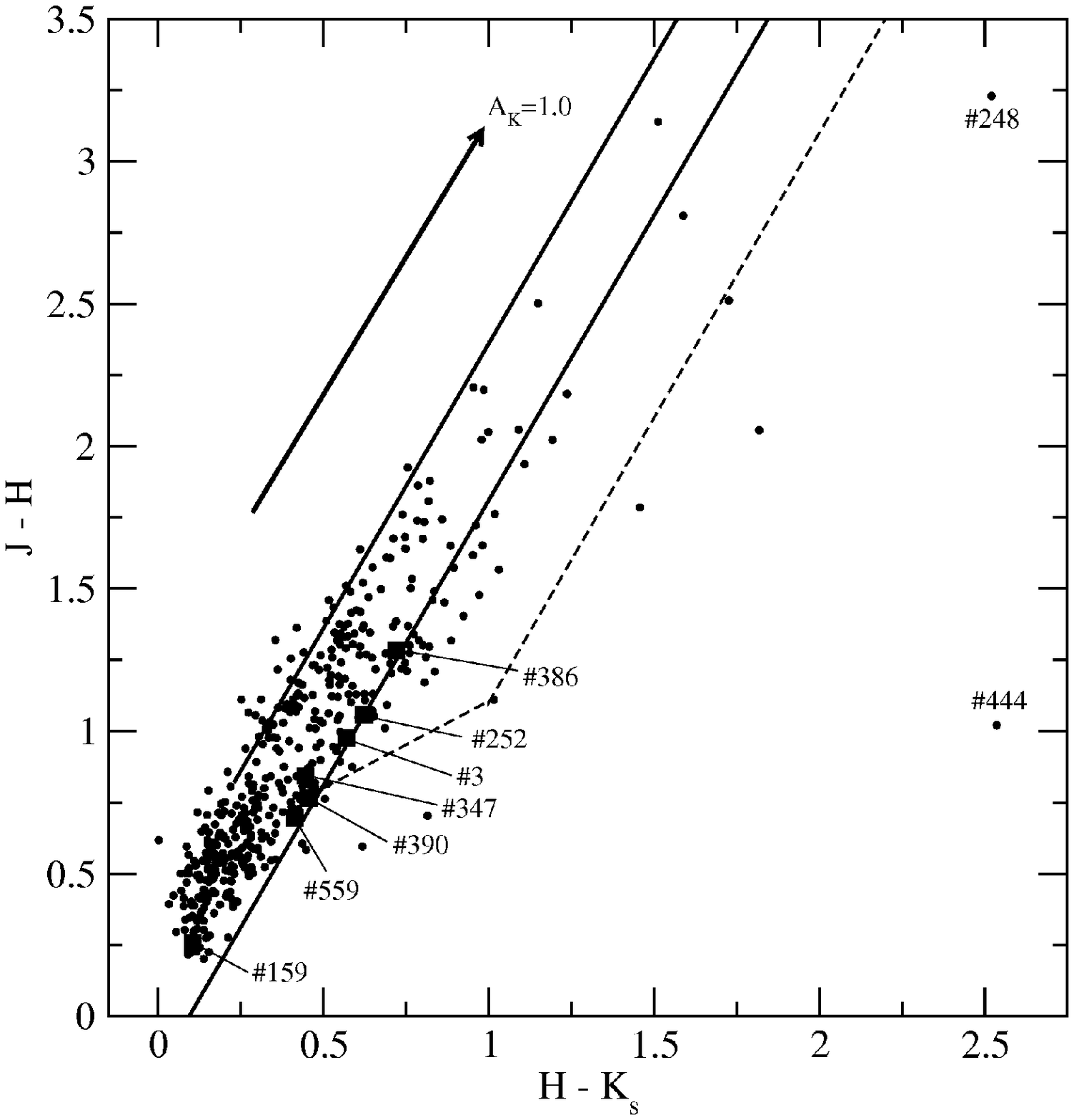}
	\end{minipage}

\caption{{Color-magnitude (C-M) and color-color (C-C) diagrams of the H\,{\sc{ii}} region W3. Triangles correspond to stars observed spectroscopically. The reddening vectors are based on \citet{Stead2009} reddening law. In the C-C diagram, the upper reddening line corresponds to M-type stars, the middle line for O-type and the dashed line delineates the zone for classical T-Tauri stars. Objects to the right (including \#248 and \#444) have a NIR excess and likely are embedded YSOs. The sequence of squares on the C-M plot indicates the position of Main Sequence stars, shifted to the kinematic distance of 4.2 kpc, which is apparently wrong by the considerable number of objects earlier than O3 stars.}}
\label{fig:diagrams}
\end{figure*}

The $J-H$ versus $H-K_{s}$ color-color (C-C) diagram is shown in Figure~\ref{fig:diagrams}b. In that diagram, the solid lines, from top to bottom, indicate interstellar reddening for main sequence M-type \citep{Frogel1978} and O-type \citep{Koornneef1983} stars while the dashed line shows the T-Tauri stars \citep{Meyer1997}. Stars located next to the O-type stars reddening line are good naked photosphere OB star candidates. Several objects are seen on the classical T-Tauri region and just a few sources are located to the right, in the young stellar objects (YSOs) domain. The reddest sources on the plot are the \#248 and \#444 objects. These sources are located on the YSO region on the C-C diagram indicating strong excess emission. Both objects also present bright $K_{s}$ magnitudes, as indicated in the C-M diagram. 

The $K_{s}$ versus $H-K_{s}$ color-magnitude (C-M) diagram is displayed in Figure~\ref{fig:diagrams}a. Squares indicate the position of the theoretical ZAMS, shifted to the kinematic distance of 4.2 kpc. The OB star candidates, identified previously on the C-C plot, are the brightest sources compared to the other stars in the W3 field-of-view.  All selected targets are limited by the $H-K_{s}$ range from 0.16 to 0.77. The \#159 source is the brightest source in the $K_{s}$-band, and also has the bluest $H-K_{s}$ color of the selected targets. This source is located in the central cluster, as shown on 2MASS and Spitzer images. This cluster is less reddened compared to the other W3 sub-regions, indicating that almost all the gas has been dissipated by the luminous stars.

The selected targets (\#3, \#159, \#252, \#347, \#386, \#390 and \#559) are labelled on both the C-M and C-C plot. 
Spectra of these stars were obtained with NIFS. Only data with photometric uncertainty less than 0.05 mag were included on the plots.

\subsection{Reddening}

An interstellar extinction law in the near infrared domain is well fit by $A_{\lambda}$ $\propto$ $\lambda^{-\alpha}$. That is, in the NIR, the extinction can be assumed as a power law. There are several interstellar extinction laws in the literature and two ``extremes'' laws were adopted: given by \citet[][$\alpha$ = 1.70]{Mathis1990} and \citet[][$\alpha$ = 1.99]{Stead2009}. These laws are extreme situations and should cover the range of reddening expected in the Galaxy. The extinction by the ISM is less effective in the Mathis law than in the case of Stead \& Hoare. 

The Mathis reddening law is given by A$_{K}$ $\approx$ 1.79$\times(E_{H-K_{s}})$ while Stead \& Hoare law is A$_{K}$ $\approx$ 1.33$\times(E_{H-K_{s}})$. An average intrinsic color ($H-K_{s}$)$_{0}$ = -0.05 for OB stars \citep{Koornneef1983} was used. The observed stars (except $\#159$ - a member of the central cluster) have an average observed color of $H-K_{s}$ = 0.5, corresponding to A$_{K_{s}} = 0.94$ mag, using \citet[][]{Mathis1990}, and A$_{K} = 0.70$ mag using \citet[][]{Stead2009}. As we are dealing with extreme interstellar extinction laws, others laws as \citet{Nishiyama2008} and \citet{Indebetouw2005}, shall give results between the ones obtained using the selected laws (Mathis and Stead \& Hoare).

There's no correlation between the total to selective extinction parameter $R_V$ in the NIR domain. Since the reddening $A_K$ is a function of the $H-K_{s}$ color -- extracted from the literature --, we cannot obtain a reasonable value of $R_V$, because NIR collapse all extinction laws, that is, there is no dependence to extinction of wavelengths longer than the size of interstellar grains with the grains themselves.

\subsection{Analysis of Spectra}	\label{respec}		

The observed spectra were compared to $K-$band spectroscopic standards presented by \citet{Hanson1996,Hanson2005}, using the classification scheme for the OB stars as defined in section \ref{reduction}.

The spectra have been divided by a low-order fit to the continuum after correction for telluric absorption. A smoothing correction (boxcar averaging of 3 neighbor pixels) was applied to the spectra to enable the detection of weak features, increasing the signal-to-noise ratio. The achieved mean signal-to-noise ratio was S/N$\sim$300. This correction allowed the identification of He\,{\sc{i}} line at 2.058~$\mu$m for objects \#159 and \#559, and the C\,{\sc{iv}} triplet for \#559.

Our data clearly show two main groups of stars: i) two stars are well classified as O-type stars by the presence of photospheric features in their spectrum; ii) four sources were identified as early B-type stars by the absence of any photospheric feature beyond the Br-$\gamma$ and Helium at 2.1126 $\mu$m and they will be discussed later as supplementary data points for the distance estimation to the W3 complex.  A single target displayed circumstellar emission in He\,{\sc{i}} at 2.058 $\mu$m and Br-$\gamma$, indicating a B-type star evolved off the Main Sequence.

\subsubsection{O-type Stars}

\begin{figure*}[!ht]
\begin{center}
\includegraphics[width=16cm]{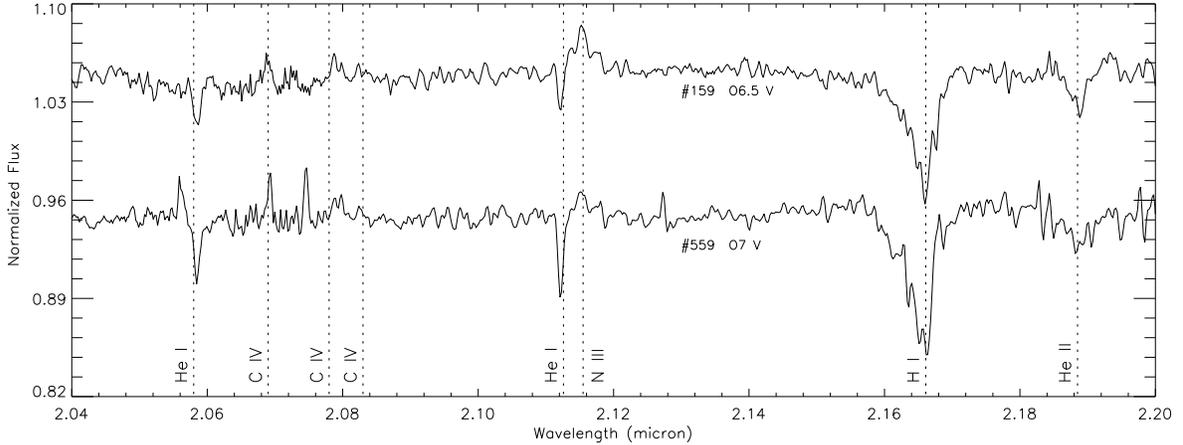}
\caption{{Spectra of O-type stars, \#159 and \#559. Lines of C\,{\sc{iv}}, He\,{\sc{ii}} and N\,{\sc{iii}} identify these sources as mid to late O-type.}}
\label{fig:o-spec}
\end{center}
\end{figure*}

The spectrum of \#159 is presented in Figure~\ref{fig:o-spec}. The presence of N\,{\sc{iii}} in emission at 2.116 $\mu$m indicates that this is an O-type star, with spectral type earlier than an O8 star.
Weak emission lines were detected at 2.069, 2.078 and 2.083 $\mu$m, which are attributed to the C\,{\sc{iv}} triplet lines, placing this star between O5-O7 type. The absorption of He\,{\sc{ii}} at 2.189 $\mu$m indicates that the spectral type is earlier than O7. 
Absorption of He\,{\sc{i}} at the blue wing of N\,{\sc{iii}} was detected. 
Both the $K$-band spectrum of HD~93130 and HD~17505, presented by \citet{Hanson1997}, seem to fit the spectra of source \#159. \citet{Patriarchi2001} have classified HD~93130 as O6\,III(f) star, while HD~17505 was classified as O6\,V by \citet{Thompson2004} and received the O6.5\,V classification by \citet{Brown2005}. 
Based on the observed features of \#159 and previous work for the classification spectra, we classify this \#159 as an O6.5\,V star.

Source \#559 (Figure~\ref{fig:o-spec}) exhibits both N\,{\sc{iii}}, in emission; He\,{\sc{i}} and He\,{\sc{ii}}, in absorption, at 2.116 $\mu$m and 2.189 $\mu$m, respectively. These features are weaker than in \#159, indicating that its spectral type is later than O6.5.
The presence of narrow emission lines at 2.069 and 2.079 $\mu$m can be associated to the C\,{\sc{iv}} triplet. These lines are relatively stronger than observed on the previous \#159 target.
The \#559 spectrum is similar to HD~190864 and HD~93222 as presented by \citet{Hanson1997}. HD~190864 source was studied by \citet{Repolust2004}, who classified this object as an O6.5\,III(f) star. \citet{Jensen2007} have adopted the O7\,III(f) classification for HD~93222. In addition, this source is located in a bright nebular region on the W3 complex, excluding a subgiant luminosity. Based on these features and classification spectra, we classify \#559 as O7\,V.

\subsubsection{B-type Stars}

The remaining targets exhibit only the Br-$\gamma$ and He\,{\sc{i}} features and their classification is limited when compared to an O-type classification scheme. 

\begin{figure*}[!ht]
\begin{center}
\includegraphics[width=16cm]{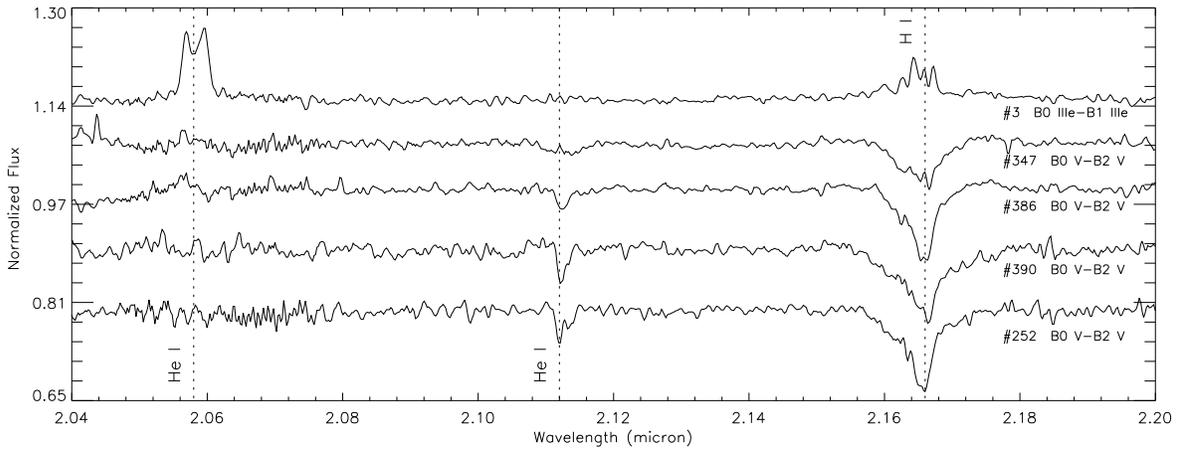} 
\caption{{Spectra of B-type stars. Typical photospheric lines of H and He are shown. The object \#3 shows circumstellar emission and should be more evolved than the other stars.}}
\label{fig:b2-spec}
\end{center}
\end{figure*}

The \#3 source (Figure~\ref{fig:b2-spec}) exhibits strong emission features. A double peaked He\,{\sc{i}} in emission at 2.058 $\mu$m, He\,{\sc{i}} triplet at 2.161 $\mu$m, 2.162 $\mu$m and 2.164 $\mu$m and, also, Br$\gamma$ at 2.166 $\mu$m are in emission. These features are commonly seen in evolved early-B stars surrounded by circumstellar disk and the adopted classification for this source ranges between B0\,IIIe and B1\,IIIe by comparison to \citet{Hanson1996} catalogue.

The remaining spectra have poor telluric correction in the region of the He\,{\sc{i}} 2.058 $\mu$m line. Objects \#347 and \#390 may have residual nebular or circumstellar He emission consistent with apparent Br$\gamma$ emission seen superposed on the broad (stellar) Br$\gamma$ absorption.
The He\,{\sc{i}} 2.112 $\mu$m absorption increases from \#347 to \#386, \#252 and reaches its maximum in the \#390 source. 
He\,{\sc{i}} and Br$\gamma$ relative absorption strengths for all four sources are compatible with B0\,V - B2\,V types. For stars earlier than B0\,V, the He\,{\sc{ii}} 2.189 $\mu$m absorption becomes more significant while types later than B2\,V do not display the He\,{\sc{i}} feature.

\begin{table*}
\begin{center}
\caption{Properties of the OB Stars in the W3 complex.}
\begin{tabular}{cccccccccc}
\hline\hline	\\ [-1.5ex]
         &				& 				&		&	&\multicolumn{2}{c}{\citet{Mathis1990}}&\multicolumn{2}{c}{\citet{Stead2009}}&              \\[4pt]
\multirow{2}{*}{Obj. ID}& R.A.& Decl.  &K$_{s}$&$H-K_{s}$&A$_{K_{s}}^{a}$&	D$^{b}$		&A$_{K_{s}}^{a}$& D$^{b}$     	& \multirow{2}{*}{Spectral type}	\\
   		 & (J2000) 		& (J2000)		& (mag)	& (mag)	& (mag)	&		(kpc)			& (mag)	&		(kpc)			&              \\[4pt]
\hline	\\ [-1.5ex]
\#159	 & 02 26 34.41 	& +62 00 42.4 	& 7.61 	& 0.11 	& 0.19 	&2.03$^{+0.74}_{-0.54}$ & 0.14 	&2.08$^{+0.75}_{-0.55}$ & O6.5\,V	\\[4pt]
\#559	 & 02 25 28.11	& +62 05 39.6 	& 8.49	& 0.41	& 0.74 	&2.24$^{+0.81}_{-0.60}$ & 0.55 	&2.44$^{+0.89}_{-0.65}$ & O7\,V		\\[6pt]
Mean$^{c}$& 			& 				&      	& 	  	&     	&2.13$^{+0.78}_{-0.68}$ &      	&2.26$^{+0.82}_{-0.60}$ &			\\[6pt]
\hline \\ [-1.5ex]
\#252    & 	02 27 21.28 & +61 54 57.2 	& 9.99 	& 0.62	& 1.12	&2.21$^{+1.19}_{-0.87}$ & 0.83 	&2.52$^{+1.36}_{-0.99}$ & B0\,V - B2\,V	\\[4pt]
\#347    & 02 25 44.86 	& +62 03 41.4 	& 8.90 	& 0.45	& 0.80 	&1.55$^{+0.83}_{-0.61}$ & 0.59 	&1.70$^{+0.91}_{-0.67}$ & B0\,V - B2\,V	\\[4pt]
\#386    & 02 25 37.51 	& +62 05 24.5  	&10.36 	& 0.72	& 1.29	&2.42$^{+1.29}_{-0.95}$ & 0.96 	&2.81$^{+1.51}_{-1.11}$ & B0\,V - B2\,V	\\[4pt]
\#390    & 02 25 27.38 	& +62 03 43.2 	& 9.99 	& 0.46	& 0.82 	&2.54$^{+1.36}_{-1.00}$ & 0.61 	&2.80$^{+1.50}_{-1.10}$ & B0\,V - B2\,V	\\[6pt]

Mean$^{d}$ & 			& 				&    	&      	&       &2.16$^{+1.07}_{-0.78}$ &      	&2.39$^{+1.19}_{-0.87}$ &		\\[6pt]

\#3      & 02 27 35.00 	& +61 55 57.2 	& 8.82	& 0.57	& 1.02	&            		     & 0.76	&               	 & B0 IIIe-B1 IIIe	\\[6pt]

\hline\hline
\end{tabular}
\end{center}
\footnotesize{
\textbf{Notes:} \\		

$^{a}$ The reddening for each star were determined from $H-K_{s}$, using $(H-K_{s})_0=-0.05$. The A$_{K_{s}}$ values were derived using the extinction laws of \citet{Mathis1990} and \citet{Stead2009}, respectively; the magnitudes were obtained from the 2MASS catalogue.

$^{b}$ The distances (D) were obtained for objects classified as Main Sequence stars and considering each reddening law.

$^{c}$ The mean value was obtained by considering only the well-classificated O-type stars.

$^{d}$ The mean distance were calculated over all OB stars. The uncertainty quoted is due to error propagation of the individual distances in both mean distances.
\\
}
\label{table:t_results}
\end{table*}

\section{Distance Determination}	\label{distance}
 
In the previous section we classified the spectra of seven stars in the W3 complex as belonging to OB-type stars. The distance to W3 was estimated using the spectroscopic and photometric results.

The M$_{K}$ of O-type stars were obtained using M$_{V}$ from \citet{Vacca1996}. For the B-type stars, we used M$_{V}$ from \citet{Wegner2007}. For both spectral types, $V-K$ values were obtained from \citet{Koornneef1983}. These photometric data were converted to the 2MASS photometric system by transformations derived by \citet{Carpenter2001}. The $K_{s}$ magnitude and intrinsic color ($H-K_{s}$) of each target were obtained from the 2MASS catalogue. 

The estimated distances are shown in Table~\ref{table:t_results}. For the derived spectral types and considering only the well classified O-type stars (\#159 and \#559), we obtain a mean distance of $2.26^{+0.82}_{-0.60}$~kpc using \citet{Stead2009}, $2.13^{+0.78}_{-0.68}$~kpc using \citet{Mathis1990} reddening law and a mean distance of $2.20^{+0.80}_{-0.64}$~kpc combining both results. Including the B stars, the distance goes to $2.39^{+1.19}_{-0.87}$~kpc  and $2.16^{+1.07}_{-0.78}$~kpc using extinction laws from \citet{Stead2009} and \citet{Mathis1990}, respectively; and $2.29^{+1.13}_{-0.83}$~kpc combining both results. The adopted uncertainty in each individual distance is given by the square root of the photometric error from 2MASS catalog ($<$ 0.05 mag) added in quadrature to the uncertainty from the absolute magnitudes ($\pm 0.67$~mag). These errors were used to derive non-symmetric errors of each distance. For the mean distance, the error was calculated by the square root of the quadratic sum of the individual errors divided by the number of stars considered.

Both O- and OB-type distances match the previous results from non-kinematic methods. 
Our result is half the value of 4.2 kpc derived by \citet{Russeil2003}, which is based on the radio recombination lines and a kinematic model \citep{Brand1993}.

\citet{Smith1978} estimated the Lyman continuum luminosity of G133.7+1.2, which is associated with an an aperture 1.6 arcmin diameter near object \#248, to be 0.88 $\times$ 10$^{50}$ s$^{-1}$ assuming a kinematic distance of 3.1 kpc.

Adopting our mean value of 2.20~kpc, as indicated by the spectrophotometric result, considerably reduces the expected ionizing flux from the radio continuum measurements of \citet{Smith1978} to $0.44\,\times\,10^{50}\,$s$^{-1}$, equivalent to the radiation of $\approx$ four O7\,V stars. The stars classified in this work gives $0.43 \times 10^{-50}\,$s$^{-1}$, which, by using \citet{Vacca1996}, translates into $\approx\,3$ O7\,V stars. It suggests there is at least one additional mid-late O-type star in the complex. This is probable, since we did not take spectra from all the bright objects in the area. The missing O-type star(s) may even be not so bright, if seated behind obscuring nebular patches.


\section{Discussion and Conclusions}	\label{discussion}

Spectra were classified based on the \citet{Hanson1996,Hanson2005} catalogues, resulting in two O-type stars, which were used to derive a mean spectrophotometric distance of $2.20^{+0.80}_{-0.64}$~kpc to these targets. 
The major source of uncertainty of the spectrophotometric method in the {\it K} band is due to the flux calibration of the absolute magnitudes \citep[$\pm0.67$~mag for O-type stars,][]{Vacca1996}, which dominates uncertainties. 

We considered the B-type stars separately from the O-type stars, because of their less accurate classification; however, their distances are in good agreement (D = $2.29^{+1.13}_{-0.83}$~kpc) with the O-type stars. The $K-$band classification of B-type stars is more uncertain due to the lack of higher ionization photospheric features; the classification is based only on He\,{\sc{i}} and Br$\gamma$ line strengths. Thus, in addition to the photometric error, a classification uncertainty is also included for these objects.
On the other hand, since the B-type stars have longer lifetimes in the main sequence stage than O-type ones, the adopted luminosity classification range puts them on or near the main sequence in young star forming regions with more confidence.

Our result is in agreement with previous trigonometric parallax distances \citep[e.g.][]{Xu2006,Imai2000,Chen2006,Hachisuka2006}. \citet[][]{Xu2006} derived the distance of 1.95 kpc for the star-forming W3(OH) region which is close in projection to the stars \#3 and \#252 studied here. The $K-$band distance to the W3 complex is also consistent with the spectrophotometric result in the optical domain \citep[][]{Humphreys1978}. The agreement between both results shows that the near infrared method is equivalent to the optical one.

Considering the recombination line velocity of -44.6 $\pm$ 5 km~s$^{-1}$, \citet{Russeil2003} found the kinematic distance of W3 as 4.2~kpc. The velocity error translates into a distance uncertainty of $\pm$ 0.6 kpc to its value. The upper limit for the spectrophotmetric distance is 3.0 kpc (using only the well classified O-type stars and the average between the extreme reddening laws). The lower limit for the kinematic distance is 3.6 kpc, in order that there is no overlap between the measurements obtained by these two techniques. 

The discrepancy between the kinematic distance to W3, which is roughly twice as large as the distance for the other methods, may be attributed to local motions of the gas deviating from the Galactic rotation.
Velocity anomalies and peculiar motion of the Perseus arm have been detected by \cite{Brand1993}. Furthermore, outward expansion of this region was shown by \citet{Heyer1998}. Also, \citet{Russeil2003} noted a velocity discrepancy of the Perseus arms of 21 km~s$^{-1}$,  which is significant as compared to typical values measured for other spiral arms ($\sim$ 3 km~s$^{-1}$).

Consideration of the C-M diagram (Figure~\ref{fig:diagrams}) confirms that the kinematic distance to W3 is too large. The brightest stars of the W3 complex are beyond the top of the main sequence line (represented as squares on the left side of the diagram) shifted to the kinematic distance of 4.2 kpc. Our results and those obtained by radio parallax suggest that the kinematic measurements require confirmation of distances by alternative methods.

Furthermore, there is a mixing of different evolutionary stages along the W3 complex, as pointed by the adopted classification of the OB stars. Recent star forming activity appears to the NE region, as indicated by nebular emission (Figure~\ref{fig:images-color}b) and also at the SW bottom, where is a maser source (W3OH). The 2 O-type and the 4 B-type stars are close to the main sequence, indicating a subsequent star formation episode.
The Be star (object \#3) looks to be evolved away from the main sequence, indicating an even older star formation activity. The separation of the double peaked line profile of He\,{\sc{i}} 2.058 $\mu$m in emission in this spectrum is 346~km\,s$^{-1}$, which is compatible with an excretion gaseous disk in a high rotation Be star \citep{Steele99}, since there is no CO emission at 2.3 $\mu$m, what is common in YSOs. 
The weak excess emission toward longer wavelengths, as shown by the C-C plot (Figure~\ref{fig:diagrams}) indicates absence of dusty environment like expected by leftovers of the accretion process. The scenario is that observed for an evolved object with an excretion disc, which produces little color changes to the photospheric light.
In this way, it seems that the star formation activity progressed from the forefront of the ISM to farther distances.

\section*{Acknowledgements}
\acknowledgements{
FN is grateful to CNPq support. He also thanks Mairan Teorodo and C\'{a}ssio L. Dal Ri Barbosa for helping with science discussion. The authors thank the useful comments and suggestions from an anonymous referee which have resulted in a much-improved version.
EF thanks L'Or\'eal-UNESCO-ABC for Brazil's 2009 For Women in Science grant. 
APM and AD are grateful to the Brazilian agency CNPq-MCT for continuous financial support. 
AD and EF acknowledge FAPESP for continuous financial support. 
PSC wishes to thank the NSF for continuous support.

These results are based on observations obtained at the Gemini North Observatory (Proposal ID GN-2008B-Q-92). The Gemini Observatory is operated by the Association of Universities for Research in Astronomy, Inc., under a cooperative agreement with the NSF on behalf the Gemini partnership: the National Science Foundation (United States), the Science and Technology Facilities Council (United Kingdom), the National Reseach Council (Canada), CONICYT (Chile), the Australian Research Council (Australia), CNPq (Brazil) and CONICET (Argentina).
}

\bibliographystyle{astron}
\bibliography{refs}

\end{document}